\RequirePackage{fix-cm}

\documentclass[sigconf,screen,nonacm]{acmart}        
\usepackage{graphicx}
\usepackage{natbib}
\usepackage{float}

\begin{document}

\title{A Computational \emph{Medical XR} Discipline}

\author{George Papagiannakis}

\affiliation{%
  \institution{FORTH - ICS Greece, University of Crete Greece, ORamaVR Switzerland}
  \city{}
  \country{}
}

\author{Walter Greenleaf}
\affiliation{%
  \institution{Stanford University USA}
  \city{}
  \country{}
}

\author{Michael Cole}
\affiliation{%
  \institution{University of Michigan Medical School USA}
  \city{}
  \country{}
}

\author{Mark Zhang}
\affiliation{%
  \institution{Brigham and Women’s Hospital Harvard University USA}
  \city{}
  \country{}
}

\author{Rabi Datta}
\affiliation{%
  \institution{University Hospital of Cologne Germany}
  \city{}
  \country{}
}

\author{Mathias Delahaye}
\affiliation{%
  \institution{Geneva University Hospitals Switzerland}
  \city{}
  \country{}
}

\author{Eleni Grigoriou}
\affiliation{%
  \institution{University of Crete Greece, ORamaVR Switzerland}
  \city{}
  \country{}
}

\author{Manos Kamarianakis}
\affiliation{%
  \institution{FORTH - ICS Greece, University of Crete Greece, ORamaVR Switzerland}
  \city{}
  \country{}
}

\author{Antonis Protopsaltis}
\affiliation{%
  \institution{University of Western Macedonia Greece, ORamaVR Switzerland}
  \city{}
  \country{}
}

\author{Philippe Bijlenga}
\affiliation{%
  \institution{Geneva University Hospitals Switzerland}
  \city{}
  \country{}
}

\author{Nadia Magnenat Thalmann}
\affiliation{%
  \institution{University of Geneva \& MIRALab SARL Switzerland}
  \city{}
  \country{}
}

\author{Eleftherios Tsiridis}
\affiliation{%
  \institution{Aristotle University of Thessaloniki Greece, Imperial College London UK}
  \city{}
  \country{}
}

\author{Eustathios Kenanidis}
\affiliation{%
  \institution{Aristotle University of Thessaloniki Greece}
  \city{}
  \country{}
}

\author{Kyriakos Vamvakidis}
\affiliation{%
  \institution{Henry Dunant Hospital Greece}
  \city{}
  \country{}
}

\author{Ioannis Koutelidakis}
\affiliation{%
  \institution{Aristotle University of Thessaloniki Greece}
  \city{}
  \country{}
}

\author{Oliver A Kannape}
\affiliation{%
  \institution{Geneva University Hospitals \& MindMaze S.A. Switzerland}
  \city{}
  \country{}
}

\date{ }% The correct dates will be entered by the editor

\begin{abstract}
Computational Medical Extended Reality \emph{( CMXR )}, brings together life sciences and neuroscience with mathematics, engineering, and computer science. It unifies computational science (scientific computing) with intelligent extended reality and spatial computing for the medical field. It significantly differs from previous “Clinical XR” and "Medical XR" terms,  as it is focusing on how to integrate computational methods from neural simulation to computational geometry, computational vision and computer graphics with deep learning models to solve hard problems in medicine and neuroscience: from low-code/no-code/genAI authoring platforms to deep learning XR systems for training, planning, real-time operative navigation, therapeutics, and rehabilitation. 
\end{abstract}

\maketitle

\section{Introduction}
\begin{figure}
    \centering
    \includegraphics[width=8.3cm]{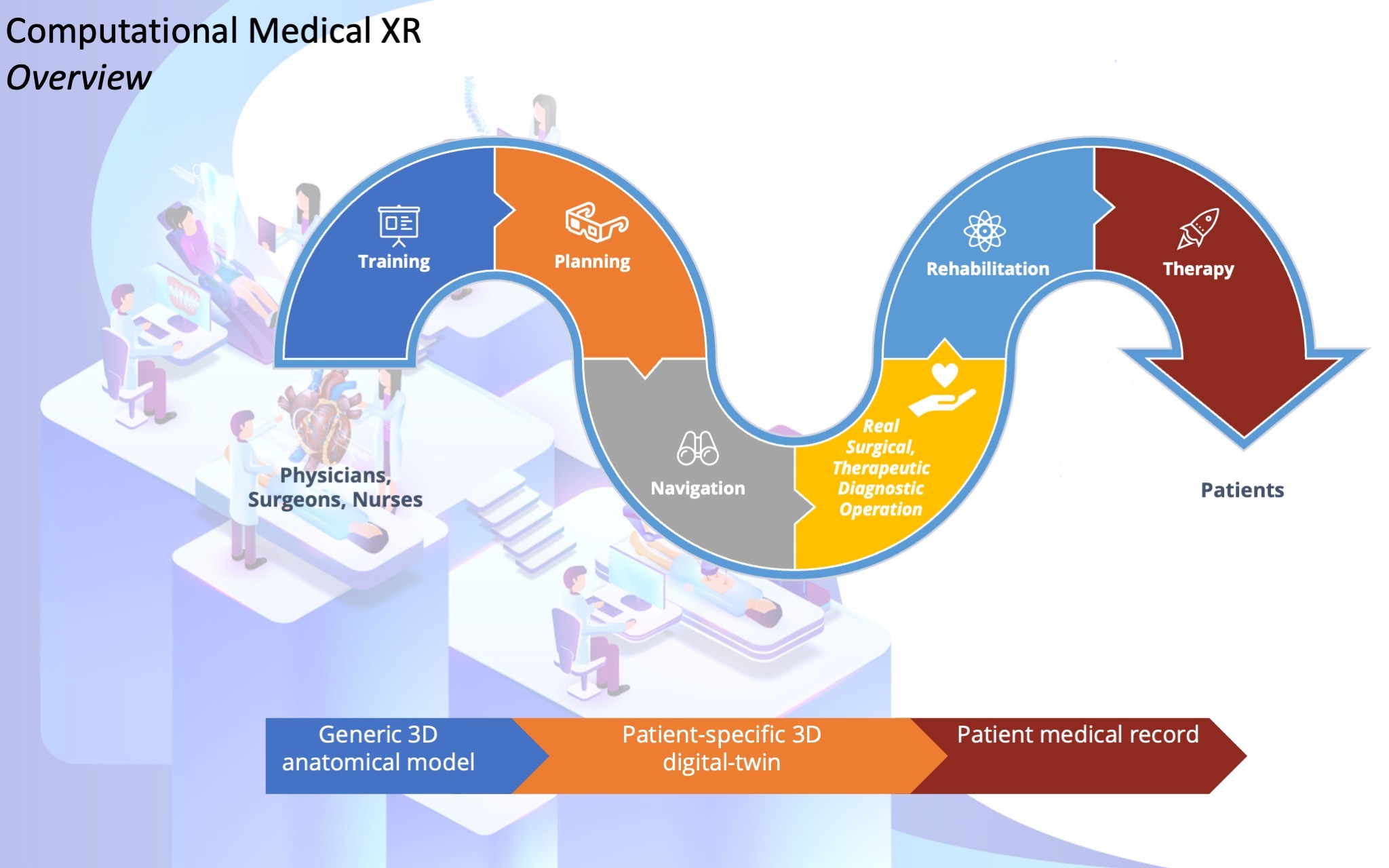}
    \caption{\cite{gp2021compXRmed} Computational Medical XR (CMXR) main application areas: Training, Planning, Navigation, Rehabilitation, Therapy \cite{frontiersUpskillResearchTopic} }
    \label{fig:CMXR_OVERVIEW}
\end{figure}

Today, 5 billion people lack access to surgical and anesthesia care, as traditional medical training methods struggle to keep up. According to OECD, over 1 billion jobs worldwide, nearly 30\% of all jobs, are likely to be transformed by technology within the next decade. In that respect, the World Health Organization predicts a shortage of 10 million healthcare professionals by 2030. Evidently, this growing need for training and continuous upskilling and reskilling of medical personnel, has become more crucial in the post pandemic era. Extended Reality (XR) coupled with spatial computing technologies emerges as a frontier in medical training, education, and empowerment, offering innovative solutions for psychomotor and cognitive skill development. In this position-survey paper, we present some of the most recent advances in the computational medical XR field, based on its definition (see figure \ref{fig:CMXR_OVERVIEW}), using state-of-the-art examples of research on simulation protocols, immersive and embodied research approaches, and steps towards more effective, user-tailored empowerment, therapy, rehabilitation, planning, navigation, upskilling and reskilling in the post-pandemic world.

\section{CMXR motivation and progress beyond the state of the art}
\label{sec:vrIntraining}

Recent CMXR related articles \cite{vrForTraining} \cite{medicalXRtaxonomy} and case study review articles \cite{10.3389/frobt.2016.00074} in industry \cite{enhanceHealthcarePostPandemic} as well as dedicated academic special journal issues \cite{frontiersUpskillResearchTopic} highlight the facilitation of Virtual, Augmented, Mixed reality (VR / AR / MR) technologies (grouped by the industry as XR) to transform and modernize the medical training model. An increasing number of published clinical trials \cite{HOOPER2019} \cite{herur2021next} \cite{bansal2022healthcare} \cite{cate2023current} measured and testified the efficacy of medical XR training and skills transfer from virtual to real. In that frame, another recent policy report \cite{XRukHealthcareReport} highlights that XR technologies can offer significant boost in experiential and collaborative learning of healthcare professionals.

XR can provide the means for remote qualitative education (knowledge) and training (skills), using affordable technology with personalized, on-demand and smooth learning curves. Based on recent major advances in the fields of 5G edge computing \cite{NeverDropTheBall2021}, neuroscience \cite{RivaEmbodiedSimulation}, MR \cite{Zikas2020VRpatterns} and spatial computing:

\begin{center}
	\emph{“VR/AR shares with our brain the same basic mechanism: embodied simulations”} \citep{RivaEmbodiedSimulation}
\end{center}

Such immersive technologies can facilitate continuous learning, provide curriculum programs and self-improvement opportunities that both expand medical professionals’ abilities as well as minimize skill gaps in their training. 

These spatial computing applications are being based on latest advances on real-time 3D graphics, computer vision, novel algebraic representations and simulation computational models. 

\begin{center}
\emph{"After more than 30 years of intense R\&D and intense validation by early adopters, medical VR technologies are now moving into mainstream"} \citep{GreenleafVRHealthcare}
\end{center}

We know from neuroscientists as well as computer scientists and educators that the feeling of Presence and the affordances of agency (body, hand interaction), Immersion and Embodiment drive neuroplasticity and long-term potentiation, which refers to the strengthening of connections between neurons in the brain based on recent patterns of activity. With each repetition, cell-to-cell signaling improves and the neural connections tighten, profoundly affecting learning, muscle memory as well as spatial memory retention. Latest advances in affordable XR hardware coupled with 5G-edge mobile and ubiquitous connectivity with massive scale/outreach are further driving this proliferation. Therapeutic XR, is also nowadays a field were the use of spatial coputing in the treatment of traumas, phobias, stress, various mental illnesses as well as in rehabilitation, is now a fact with clinically proven results \cite{GreenleafVRHealthcare,vrada2021}.

\section{CMXR in the post-pandemic era}
\label{sec:vrInpostpandemic}

Covid-19 has forever accelerated advances in this field and elevated spatial computing technologies from an optional luxury to indispensable necessity for the roughly 3000 medical schools worldwide, the 8400 medical device companies and the 600 surgical training centers, that were all challenged by covid-19 lockdowns. The integration of medical XR training in medical curricula has become a new norm confirming the recent Accenture industry report \cite{AccentureNewReality}, refered to as the “quiet revolution”, where enterprise VR vastly outnumbers consumer VR applications. Furthermore, healthcare professionals have now acknowledged that training simulators are the number one immediate application of computational XR in healthcare \cite{InsiderIntelligence}. The forefront of this revolution is the unexpected new insight that “the rising demand for virtual training across industries drives the VR market”, a \$15B market in 2020, expected to grow to \$57B, as was also recently reported \cite{GrandViewResearch}.

However, the looming question is: \emph{who is going to “XR-ify”(translate into XR) the 400+ basic medical procedures and their countless variations?}

These fundamental medical procedures \cite{MedicalProcedures} are constantly being updated and/or being extended. Moreover, the significant advancements in medical device equipment for diagnostic or therapeutic procedures, such as robotic surgery, exponentially evolves the range of variations within these basic medical procedures, necessitating advanced and comprehensive training for medical personnel, such as med-school students, residents, young or experienced surgeons or physicians of any specialty or nurse. This ongoing need for healthcare professionals to continuously upskill / reskill to adapt to the latest advancements in the field and remain compliant with regulatory agencies, along with the imperative demand for institutions to train a sufficient number of professionals to meet the escalating demand, poses a perpetual challenge.  
Traditional medical training methods either utilize cadavers, which are scarce, expensive and logistically limited, or real patients, which pose issues of limited opportunity, repeatability and possible adverse outcomes. This issue is intensified as OR (Operating Room) times are capped for trainees/residents and a decreasing number of them enter the workforce. Similar needs/problems are also witnessed in the adjacent verticals of vocational or corporate skills training.

Although the excellent de facto 3D/XR content creation industrial standards of Unity and Unreal Engines act as the essential enabling technologies and general authoring platforms, their general-purpose coverage ( from games to automotive, engineering, construction to simulation and film/ broadcasting) requires substantial efforts from highly skilled developer teams from various technological disciplines (computer graphics, deep learning, computer networking, human-computer- interaction, gamification, game development, UI/UX design, affective computing), coupled with teams of education, medicine, psychology professionals and 3D artists /designers. Consequently, the “XR-ification” of the basic medical procedures, both in extensive depth and exhaustive breadth, poses significant challenge. This challenge is not solely rooted in the lack of XR content, but predominantly arises by the lack of dedicated XR medical authoring tools and metaphors \cite{Zikas2020VRpatterns}, which are essential in accelerating the development process.

\section{The current need for CMXR}
\label{sec:compMedXR}
The unique benefits of simulation-based training in environments with multiple and complex regulations, as the healthcare industry, are already testified by aerospace and aviation case. In such cases pilots must undergo mandatory simulation training for several types of aircrafts from various aviation authorities worldwide, which facilitates increased safety in aviation worldwide. Several mandates for simulation-based training have also been put forward in medicine, and it is only a matter of time till XR simulation education labs for training and competency assessment will be mandatory on a worldwide basis as well. Medical education will be significantly boosted via experiential learning, since learning by doing has proved to be much more effective than the traditional master-apprentice model and the “see one, do one, teach one” concept.

The XR-ification process of the main medical procedures may be accelerated by leveraging new software platforms and simulation-training authoring tools \cite{Just2003, Zikas2020VRpatterns}, specifically tailored for medical XR training. These computational tools will allow rapid prototyping of medical XR training, as part of the “Medical XR” fields \cite{medicalXRtaxonomy} with special focus on computational medical education, preoperative planning and real-time operative navigation, therapy and rehabilitation under a new, holistic integrated computational XR systems approach (figure \ref{fig:CMXR_OVERVIEW}). 
\begin{figure}
    \centering
    \includegraphics[width=8.3cm]{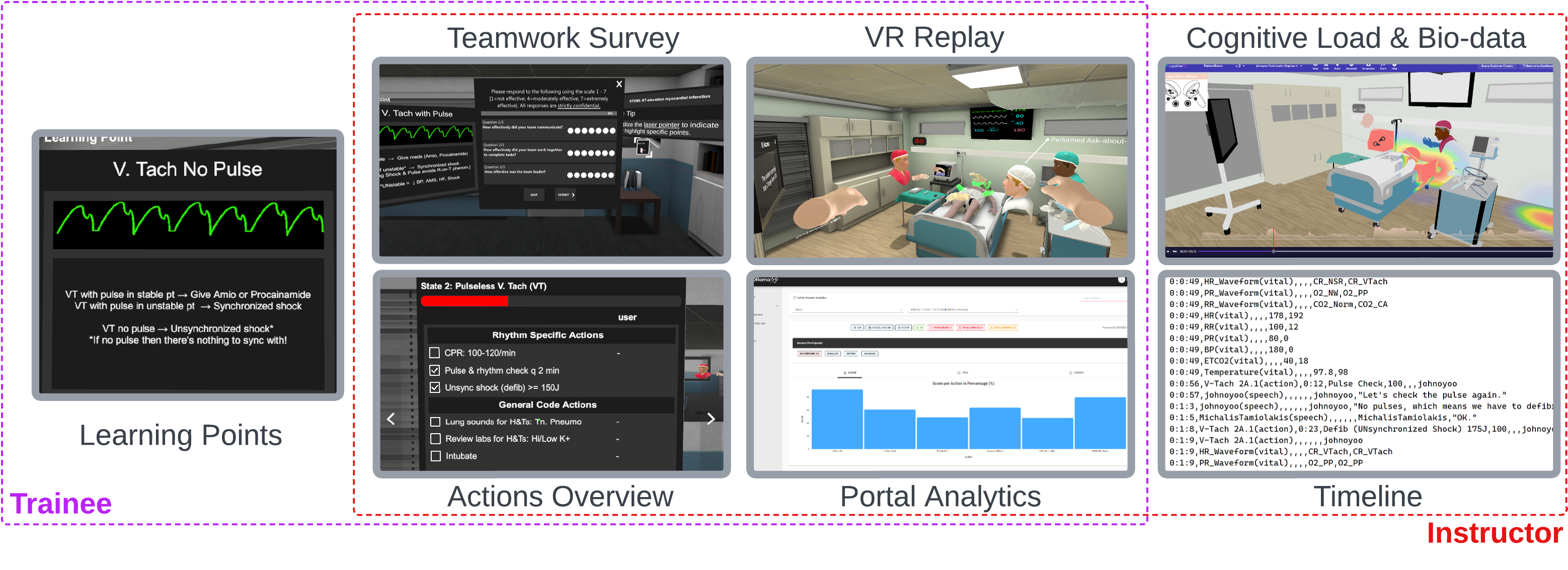}
    \caption{A set of tools that are designed to provide 
    information useful for understanding and enhancing 
    CRM during and after collaborative multiplayer medical
    XR sessions.}
    \label{fig:CRM_US}
\end{figure}

Such prototyping tools will provide medical institutions the ability to author in a low-code/no-code manner, control and develop their own XR training material, ensuring that their medical professionals are properly and continuously trained aiming in optimal patient outcomes and fewer medical errors/complications. Medical institutions should be able to rapidly author their XR medical procedures either independently or by outsourcing parts of their curricula, while retaining the ownership of the final source code. This ownership will facilitate easy updates, extensions, and modifications to their XR training content and will also drive further adoption and customization of their medical XR modules. In that respect, medical XR training authoring tools will enable subject-matter-experts create their own XR learning modules, in a similar manner they create their presentation slides today. A recent testament to this trend, is the in-house VR content creation department of the International Committee of the Red Cross and the state-of-the-art content they are creating \cite{ICRCvr}. Early scientific evidence is provided by \citep{10038619} on the real-use and skills transfer provided by such computational medical XR low-code authoring platforms, which are now been employed by leading medical organisations worldwide such as \cite{CVM}.

\section{Survey of latest use cases of Computational Medical XR (CMXR)}

\subsection{Crew Resource Management}

Crew Resource Management is vital when it comes to training medical teams, particularly in high-stress environments like surgeries or emergency situations. Ongoing studies, such as \cite{ireactPoster2024}, pioneer the integration of XR technology, specifically VR, to bolster Crew Resource Management (CRM) in medical training scenarios.

Collaborating closely with experts from the University of Michigan Medical School, the research team developed a suite of VR tools meticulously crafted to enhance teamwork, communication, and decision-making skills essential for ensuring patient safety and optimizing outcomes. By immersing trainees in realistic medical environments, these tools provide a unique opportunity for hands-on practice and assessment.

The incorporation of features such as real-time analytics visualization, cognitive load monitoring, and session replay capabilities sets a new standard for comprehensive evaluation and feedback mechanisms in medical education (see Figure~\ref{fig:CRM_US}). This innovative use of XR technology not only advances the understanding of CRM principles but also holds promise for revolutionizing medical training methodologies, paving the way for safer and more effective patient care.

\subsection{Non-linear Collaborative, Networked Simulations}

\begin{figure}
    \centering
    \includegraphics[width=8.3cm]{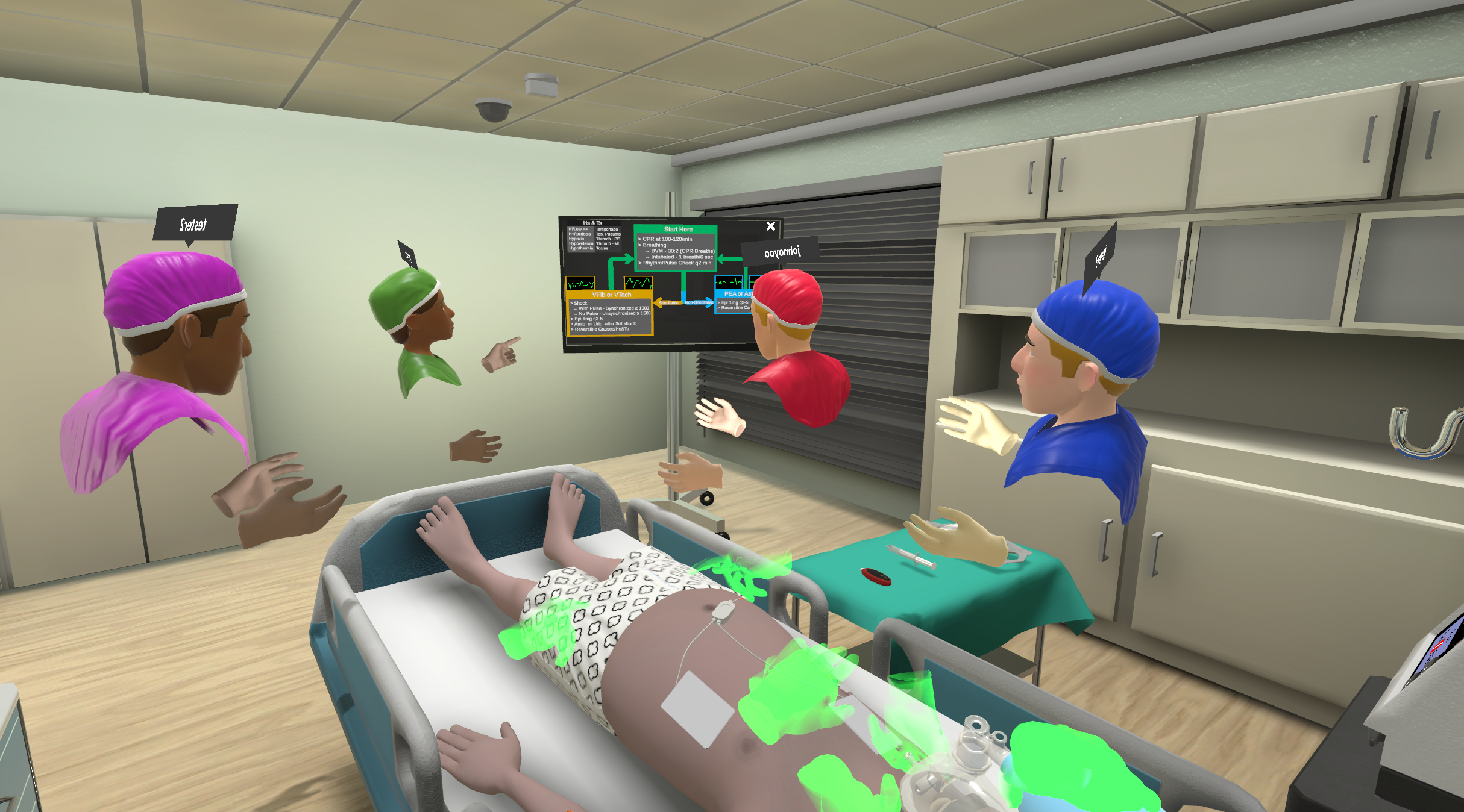}
    \caption{Medical personnel must adjust their actions and collaborate based on the patient's vital signs and previous responses during cardiac arrest resuscitation.}
    \label{fig:iREACT}
\end{figure}

Cardiac arrest resuscitation (CAR) is a critical medical procedure
that requires quick and effective intervention. Traditional Cardiopulmonary Resuscitation (CPR)
training methods often fall short in adequately preparing medical
personnel, leading to skill decay and poor survival outcomes from
cardiac arrest, primarily due to the complex techniques and equipment
involved. Recognizing this challenge, VR emerges as
a potential solution to provide an environment where complex patient
status patterns can be simulated, allowing medical teams to refine
their communication skills and enhance skill acquisition and
retention.
In this context, "iREACT" \cite{ireactPoster2024} was developed as an immersive and realistic
VR training experience for medical professionals, aiming to bridge the
gap between theory and practice in CAR training (see
Figure~\ref{fig:iREACT}). Leveraging the unique features of VR, iREACT
offers healthcare professionals an innovative, collaborative
multi-user learning tool that promotes teamwork skills and provides
individualized feedback targeting specific learning needs. It offers
an immersive experience that mirrors the high-stress environment of a
real cardiac arrest scenario while examining users' cognitive load and
focus of attention to gain insights into the cognitive and behavioral
drivers of clinical performance and medical errors.

\subsection{Surgical Robotic digital-twins}

\begin{figure*}
    \centering
    \includegraphics[width=0.9\textwidth]{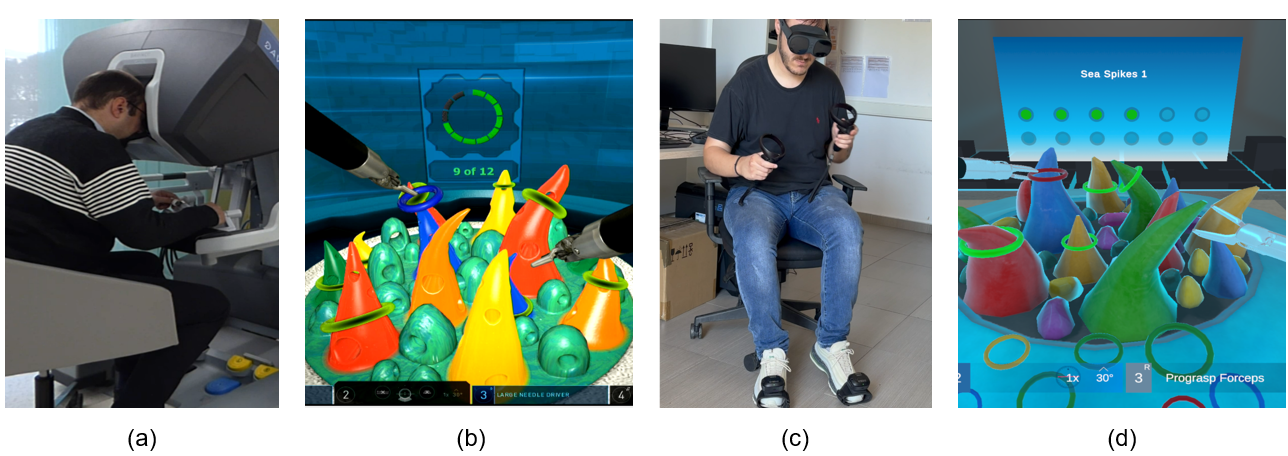}
    \caption{A comparison between a traditional SRS simulation and its digital twin, VR Isle Academy \cite{vrIsleAcademy}. Images (a) and (b) illustrate a contemporary SRS simulator, featuring a user operating from the surgeon's console. Conversely, images (c) and (d) highlight VR Isle Academy, where users manipulate a simulated SRS digital twin using an inside-out VR head-mounted display (HMD), controllers, and feet trackers}
    \label{fig:SRS_UC}
\end{figure*}

In the realm of surgical robotics, the demand for effective training
solutions to operate advanced Surgical Robotic Systems (SRS) like the
da Vinci SRS is ever-present. However, traditional training methods
are often hindered by cost constraints and limited access to SRS
equipment. Addressing this challenge, VR Isle Academy 
\cite{vrIsleAcademy} emerges as a cost-effective and immersive VR
training solution tailored for operating SRS. This portable and
device-agnostic VR simulation offers ultra-realistic experiences with
hand and feet tracking support (see Figure~\ref{fig:SRS_UC}). Users
can undergo unsupervised training sessions anytime, benefiting from
real-time error assessment and comprehensive analytics to enhance
their robot-controlling skills. Through innovative integration of VR
tracking technology and intuitive user interfaces, VR Isle Academy
promises to revolutionize SRS training, offering users an immersive
and accessible educational experience tailored for the demands of
modern surgical robotics.

\subsection{Enhanced AI-Powered Assistants}
\label{sec:IDS}
Within computational medical XR, there exists a pressing need for an
AI-powered interactive surgeon capable of training diverse individuals
– including medical students, residents, and medical personnel – in
surgical procedures, independent of the presence of a physical
surgeon. Surgeons, burdened with heavy patient loads and
administrative duties, require support. Therefore, there is a critical
demand for AI interactive surgeons to train future medical
professionals across various surgical disciplines.

\begin{figure}
    \centering
    \includegraphics[width = 7.3cm ]{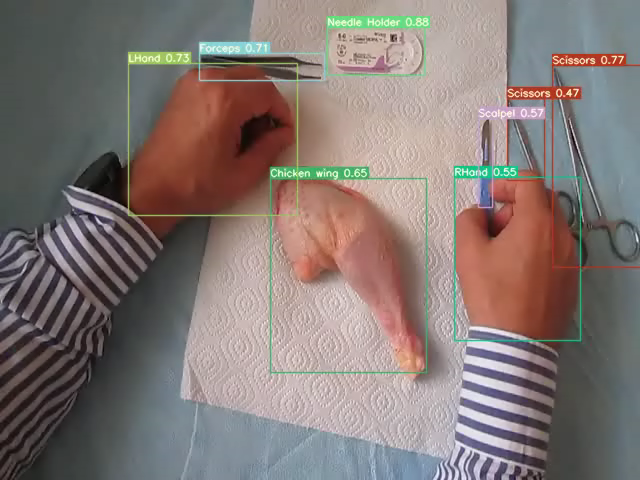}
    \caption{Creation of a skilled surgeons dataset.
    Cameras Recording hand gestures and instruments positions.}
    \label{fig:IDS_UC}
\end{figure}

As part of an ongoing project, known as the \textit{Intelligent
Digital Surgeon} (IDS) \cite{IDS}, 
researchers are developing a conversational AI
surgeon equipped to evaluate trainees' movements and gestures. To
achieve this, a dataset comprising optimal gestures used by proficient
surgeons for incisions and suturing of the skin has been defined and
annotated. The selection of these surgical procedures is informed by
the general need for medical students to perform skin surgery, often a
foundational aspect of surgical training.

A novel vision-based method \cite{transG} has been devised to analyze
students' gestures. Concurrently, a dataset of students' gestures is
being created to enable the AI digital surgeon to identify and correct
errors in their movements. A novel assessment framework will empower
the digital surgeon to provide both oral and visual guidance to
students. Specifically, Visual Large Language Models (VLLMs) are
employed to respond to trainees' queries. This enables the AI digital
surgeon to demonstrate optimal gestures while offering verbal
explanations to students (see Figure~\ref{fig:IDS_UC}).

\subsection{Training Simulation in intraoperative neuromonitoring}

\begin{figure}[H]
    \centering
    \includegraphics[width=8.3cm]{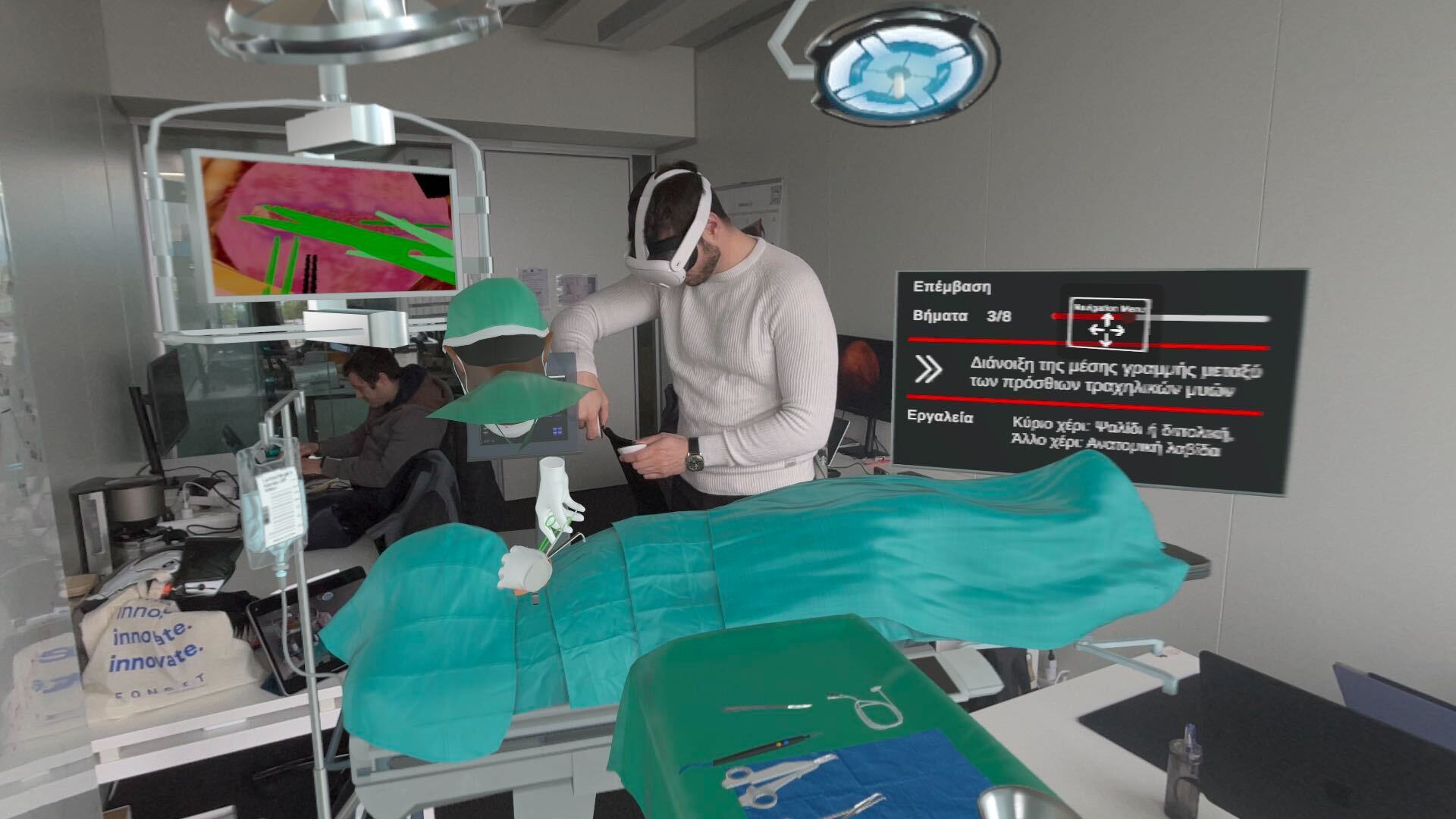}
    \includegraphics[width=8.3cm]{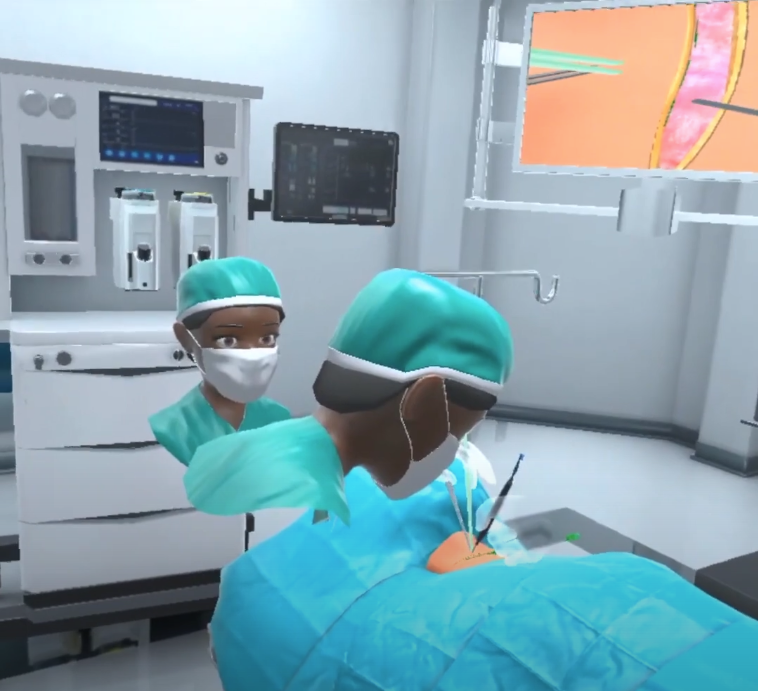}
    \caption{(Top) An AR co-op session where a user performs intra-operative neuromonitoring during a thyroidectomy, using the Apple Vision Pro. (Bottom) The respective session
    performed in VR.}
    \label{fig:HSES_UC}
\end{figure}

VR Training Simulation in intraoperative neuromonitoring stands at the forefront of endocrine surgical education, particularly focused on thyroidectomy (see Figure~\ref{fig:HSES_UC}). Utilizing the state-of-the-art Apple Vision Pro mixed reality platform, which seamlessly blends AR and VR, this simulation introduces an unprecedented approach to training. By harnessing this innovative technology, the simulation offers a unique blend of immersive VR experience, that deeply engage users in a virtual environment, and enhanced graphics through AR elements, overlaying digital information onto the real-world view, seen through the VR HMD. Trainees, using the hand tracking feature of Apple Vision Pro, can interact with the virtual environment intuitively, allowing for a more natural and immersive training experience. Primary aim of the training simulation is to provide a secure and captivating environment for learners to master the intricate nuances of electrode placement with unparalleled clarity and precision. Furthermore, the simulation offers comprehensive analytics for performance tracking, enabling learners to monitor their progress and refine their skills with precision. In case of an incomplete training step, trainees may revisit and correct their actions, fostering iterative learning and mastery. Additionally, the platform supports collaborative training, where educators can guide and instruct learners in real-time, enhancing the educational experience through personalized feedback and interaction.

This multifaceted approach ensures that trainees not only acquire the necessary skills but also cultivate confidence and competence in intraoperative neuromonitoring techniques. This groundbreaking innovation, available across multiple platforms including XR devices and mobile devices, marks a paradigm shift in endocrine surgical education. Implemented within neuromonitoring seminars at School of Medicine at Aristotle University of Thessaloniki, 
under the guidance of the Hellenic Society of Endocrine Gland Surgery, this VR application not only enhances the proficiency of medical professionals but also sets a new standard for surgical education.

\begin{figure*}[H]
    \centering
    \includegraphics[height=4cm]{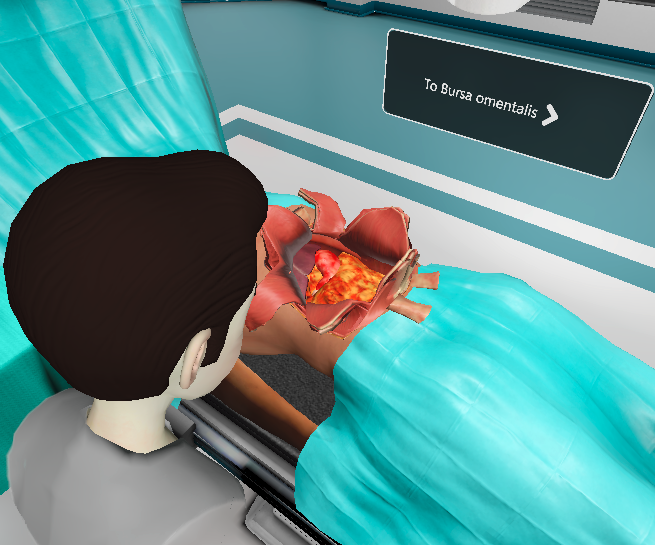}
    \includegraphics[height=4cm]{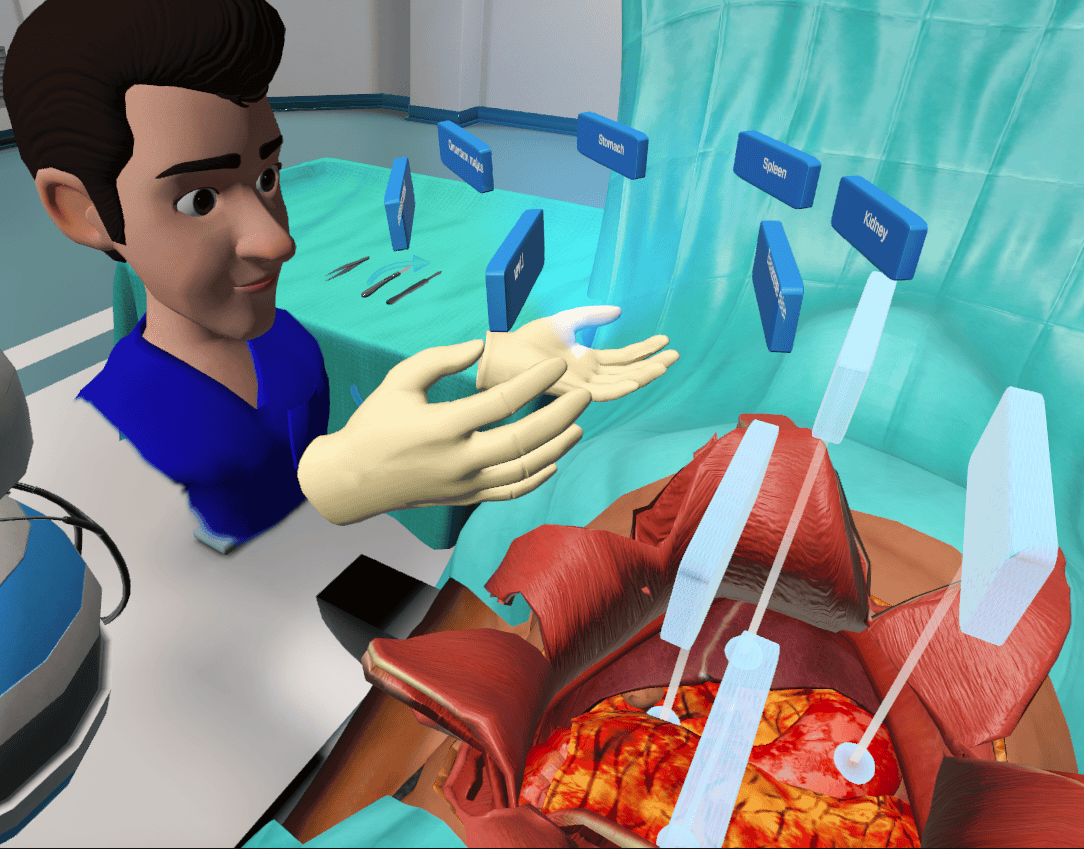}
    \includegraphics[height=4cm]{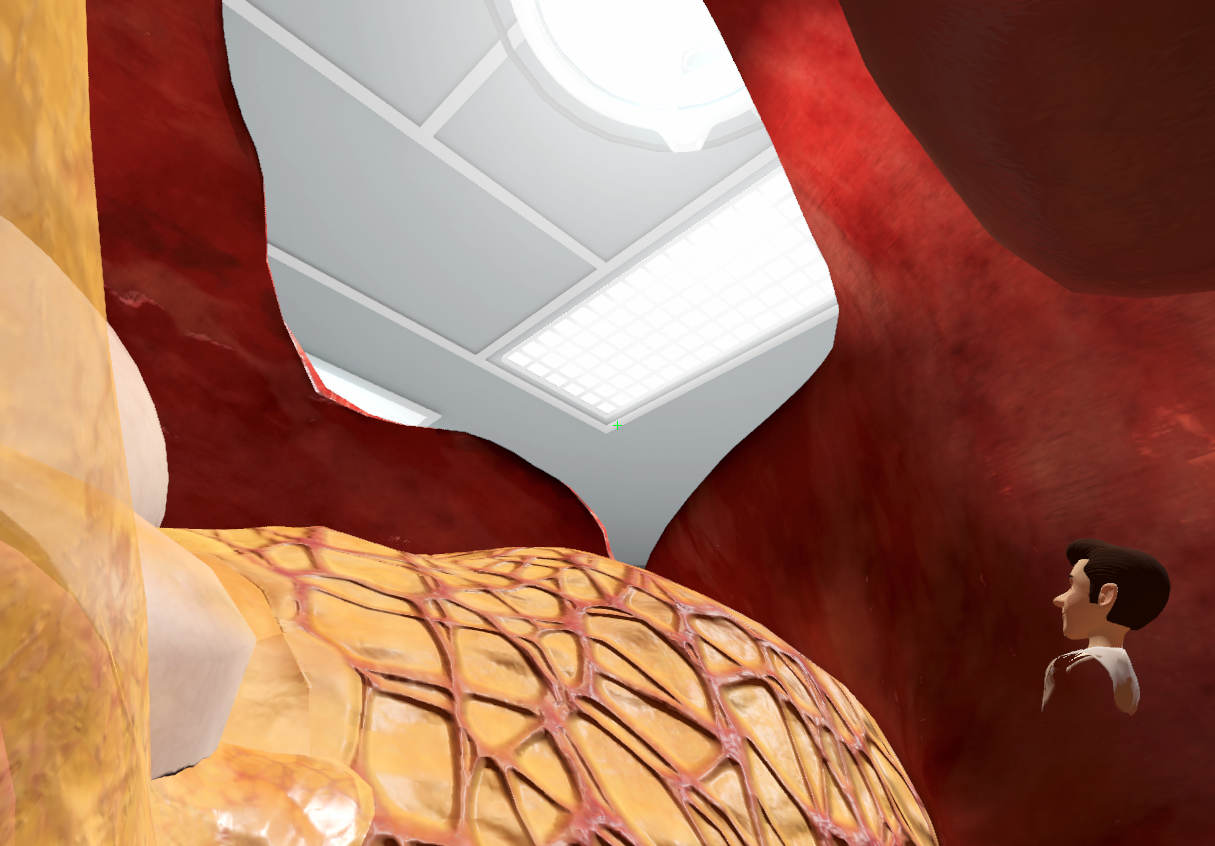}
    \caption{(Left) While performing a high-fidelity surgical operation, users have the option to explore the inner abdominal area. (Mid) Upon selection, users are "shrunk" and transported to the abdominal realm, initiating an immersive exploration journey. (Right) The inclusion of gamified elements, such as object labeling, enhances the learning impact of these experiences.}
    \label{fig:cologne_UC}
\end{figure*}

\subsection{Enhanced Storytelling with Gamified Educational Elements in CMXR}

Mastering surgical procedures requires extensive repetition to
solidify knowledge retention. Virtual simulations play a vital role in
this process by offering a platform for repeated practice. However,
achieving proficiency goes beyond merely replicating steps; it
involves enhancing psychomotor skills. Projects like the ongoing 
Topographical Anatomy by the University Hospital of Cologne \cite{cologne} go
further by immersing trainees in detailed explorations of internal
anatomy. Additionally, they employ interactive tasks such as
label-assigning to deepen understanding and reinforce learning. These
innovative approaches not only enhance knowledge acquisition but also
contribute to long-term retention, thereby revolutionizing surgical
training methodologies (see Figure~\ref{fig:cologne_UC}).

\subsection{Verified Impact of Therapeutic CMXR Platforms}

Therapeutic virtual reality  has emerged as a promising avenue for cognitive and physical training among individuals with Mild Cognitive Impairment (MCI). Nevertheless, existing VR platforms often fail to resonate with older adults experiencing cognitive decline and neglect to cater to their specific requirements.

In a clinical trial, researchers set out to develop and assess the acceptability, usability, and tolerability of VRADA, an immersive VR platform tailored for older adults with MCI symptoms. VRADA facilitates concurrent physical and cognitive training through a dual-task methodology.

\begin{figure}
    \centering
    \includegraphics[width=7.5cm]{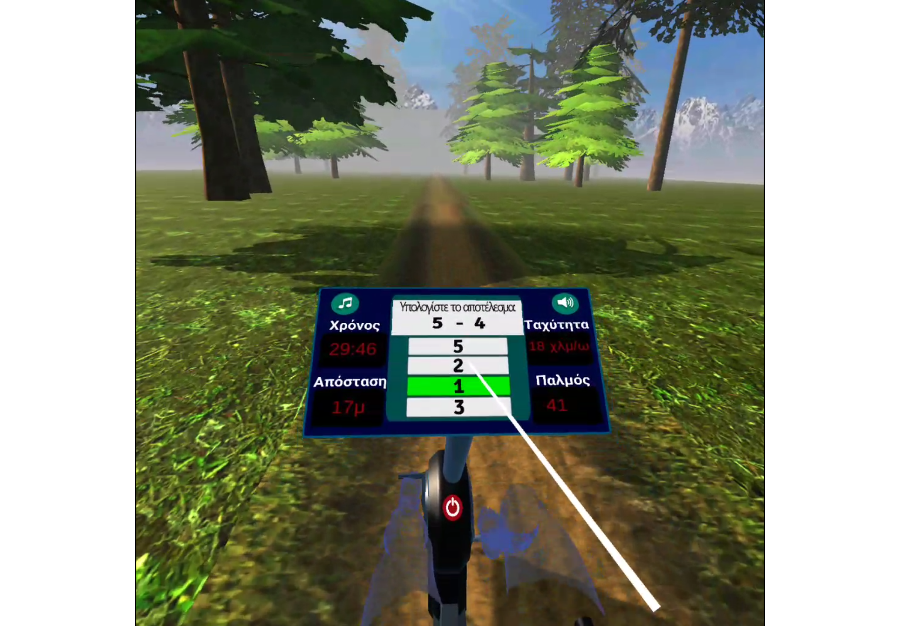}
    \caption{The VRADA application: User selects the answer to the math calculation exercise, during a cycling.}
    \label{fig:vrada}
\end{figure}

The investigation revealed that VRADA empowers users to customize their cycling duration and engage in physical exercise while concurrently tackling cognitive exercises, such as math quizzes 
(see Figure~\ref{fig:vrada}).
The application offers real-time feedback on task performance and allows users to assess their performance post-engagement. A controlled trial \cite{vrada2021} underscored VRADA's efficacy in bolstering physical and cognitive well-being among older adults with MCI symptoms, signaling its potential as a valuable therapeutic resource.

\subsection{Swift CMXR adaptation to urgent medical needs}

\begin{figure}[H]
    \centering
    \includegraphics[width=8.3cm]{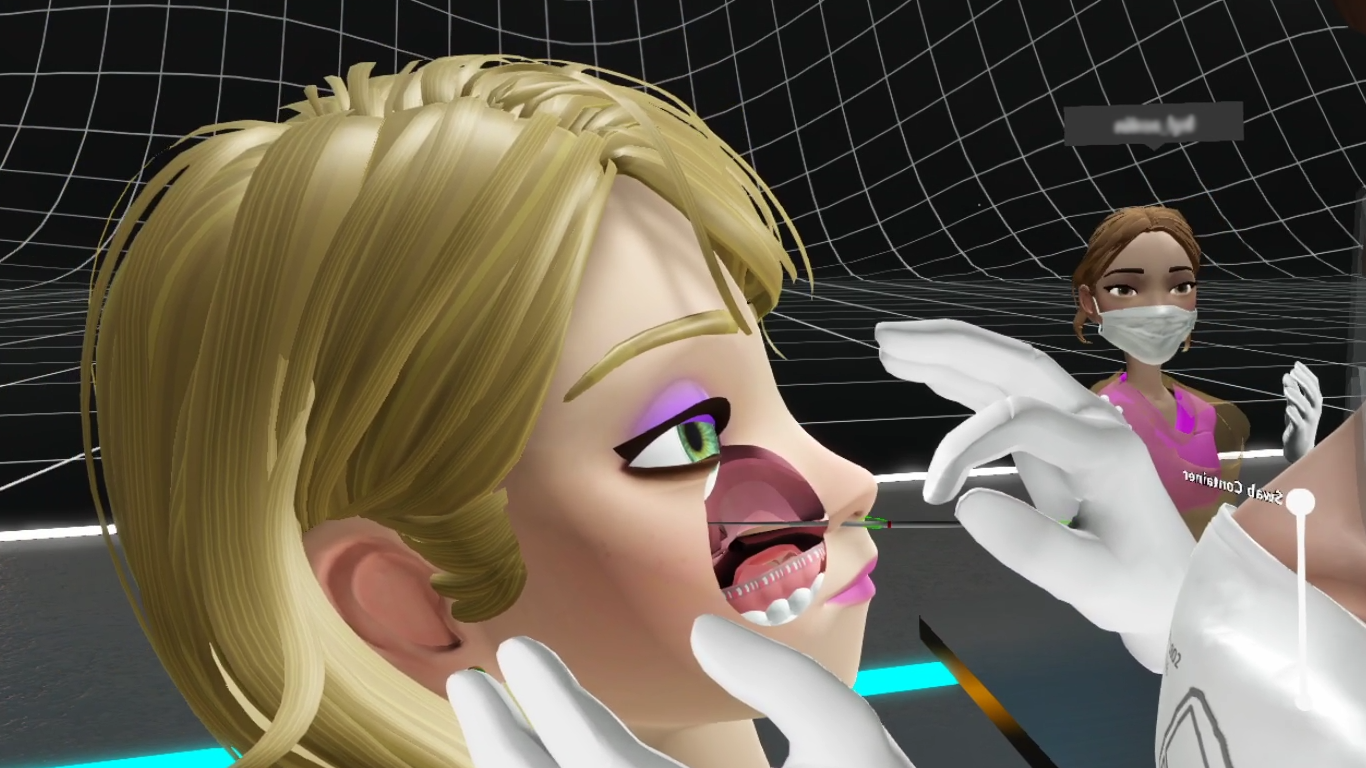}
    \includegraphics[width=8.3cm]{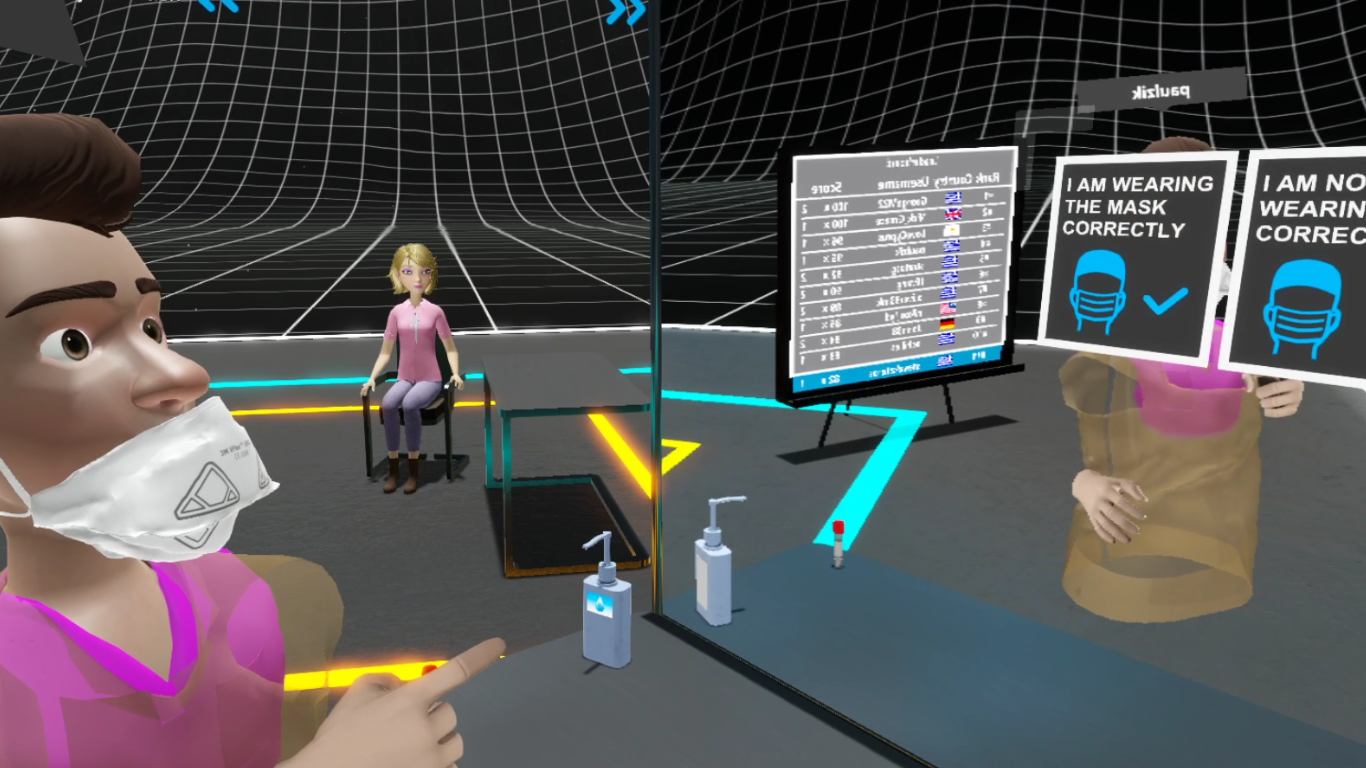}
    \caption{(Top) The instructor demonstrates the proper technique for performing the swab test. (Bottom) Cognitive questioning and the reflective mirror enhance the effectiveness of the learning process.}
    \label{fig:CVRSB_UC}
\end{figure}

During the Covid-19 pandemic, medical personnel were tasked with
rapidly training new generations while navigating the crisis. Although
the crisis has since subsided, the potential for similar situations to
arise remains. Medical XR approaches are positioned to provide relief
in such circumstances by offering tools that assist doctors in quickly
adapting to the ever-changing landscape of medical training. ``Covid-19
VR Strikes Back'' \cite{cvrsb} exemplifies this, offering 
a VR training
application focused on nasopharyngeal swabbing and Personal Protective
Equipment protocol during the pandemic (see Figure~\ref{fig:CVRSB_UC}). 
Validation through a pilot study at the Inselspital, 
University Hospital Bern, Switzerland,
demonstrated significant improvements in sensorimotor performance
among VR-trained participants compared to traditional methods.

\subsection{Virtual Objective Structured Clinical Examinations (OSCEs)} % 
\label{sec:VOSCE}
Objective Structured Clinical Exams (OSCEs) are essential for assessing medical students' practical clinical proficiency \cite{Khan2013PartI} across affective, cognitive, and psychomotor domains \cite{Bloom1986}. On the contrary, OSCEs present organizational challenges and individual burdens for examiners and students alike \cite{Bevan2019}.

Typically, an OSCE involves multiple simultaneous stations, each featuring accurately thought-out and validated clinical scenarios, executed by a trained actor-patient, 
along with an examiner. 

\begin{figure}[H]
    \centering
    \includegraphics[width=0.4\textwidth, height = 500pt]{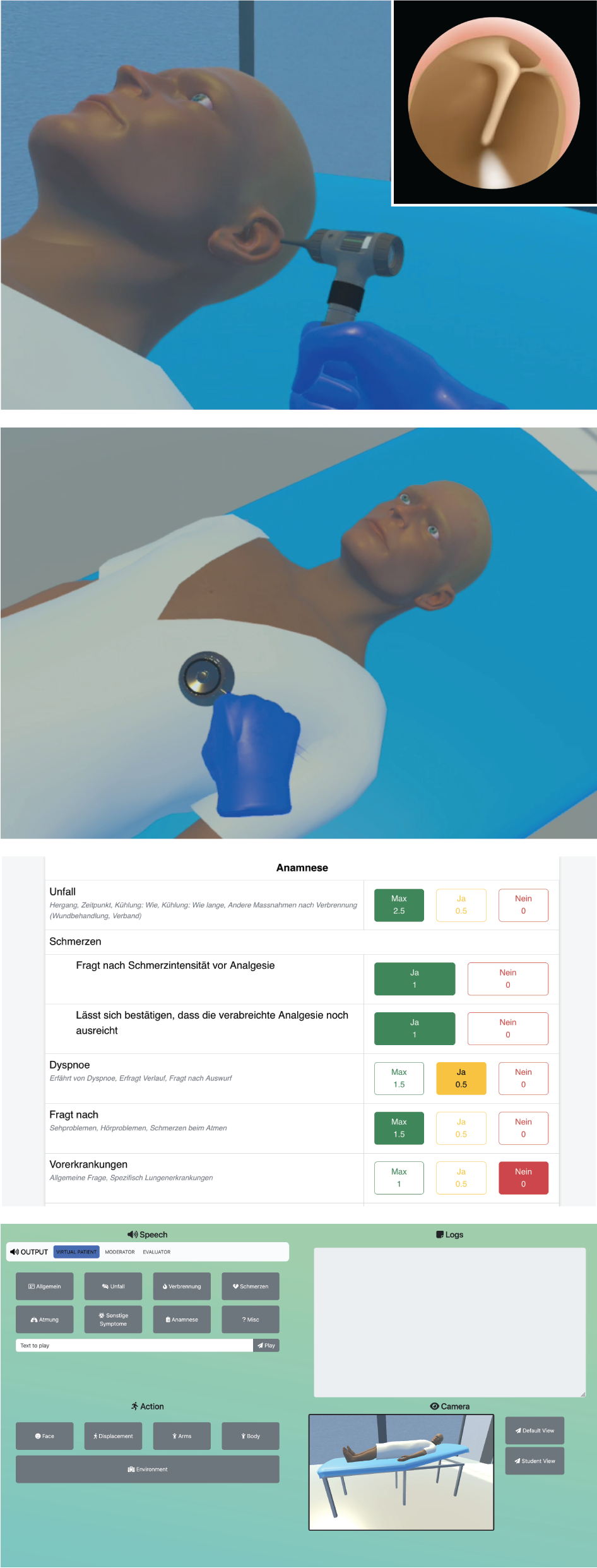}
    \caption{(Top) The virtual OSCE station allows for integration of medical images emulating the view through on otocoscope, when the device is close enough to the ear, see inset, as well as playing healthy or pathological sounds during an auscultation. The third image shows an example checklist that is automatically populated during the exam. (Bottom) The control interface for the exam moderator, with a live transcript of the discussion along with an independently controllable, real-time view of the scene.}
    \label{fig:vosce}
\end{figure}

Viva VOSCE project \cite{VOSCE} 
simulates in VR an official OSCE station, used at the medical examinations at the University of Bern, Switzerland. The creation of a virtual “simulated patient” ensures high reliability in behavior across all examined medical students. Actor fatigue or differences in interaction with examinees can be avoided, removing potential sources of biases and minimizing confounding factors. In addition, a replay functionality allows examiners to revisit key moments of the examination to moderate results. As the system provides exam transcripts, 
standard tasks
are evaluated automatically, allowing the examiner to focus on interpersonal skills. Furthermore, many aspects of the examination that cannot be delivered in traditional OSCE scenarios, without relying on the disrupting use of printed reports or supplemental images (i.e., inspection of the ear with an otoscope), were seamlessly integrated in the VR simulation (see Figure~\ref{fig:vosce}).

\subsection{CMXR for Public Protection Disaster Relief}

Emergency medicine should play a central role in Public Protection Disaster Relief (PPDR) and management due to the overwhelming number of bodily injuries that typically occur during most forms of disasters. Various types of pre-hospital emergency treatments, such as soft tissue wounds, orthopedic trauma, and abdominal surgery [1], must be provided to injured patients on-site by emergency medical teams, before their transfer to a hospital. It is well known

\noindent
 that pre-hospital treatment is a critical phase for emergency injuries. In many cases, the relatively minimal surgical knowledge of first-aid responders and the absence of medical experts at the scene result in many incidents remaining untreated until the patient reaches the hospital.

\begin{figure}
    \centering
    \includegraphics[width=6.8cm]{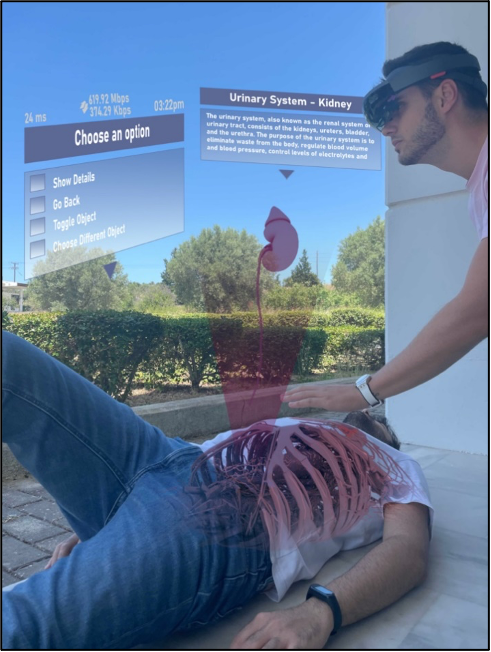}
    \caption{PPDR first responder is visualizing on-site the overlayed medical instructions and deformable objects on top of the patient}
    \label{fig:ppdr}
\end{figure}

In the context of 5G-EPICENTRE project \cite{5G-EPICENTRE}, the developed use case application ”AR-assisted emergency surgical care” \cite{kamarianakis2023ar}, provides first-aid responders, situated on a disaster site, a powerful tool that will help save lives in peril. Using untethered highly-portable AR HMDs, the PPDR responders will be able to visualise various deformable internal body parts on top of the patient effectively enabling surgeons to see inside (visualizing bones, blood vessels, etc.) and perform surgical actions following step-by-step instructions (see figure \ref{fig:ppdr}). Additionally, in the context of FIDAL project \cite{FIDAL}, the developed use case application "On-site XR-assisted emergency surgical operations", extended the previous PPDR application to enable the collaboration between on-site first-aid responders, using AR HMDs, and indoor medical experts, using VR technologies. Outdoor emergency medical surgery teams use the AR HMDs to visualize overlaid deformable medical models directly on top of the patient body and perform surgical actions following step-by-step instructions. Indoor medical experts, that visualize the scene in VR HMDs, have direct communication with the first-aid responders, providing them real-time medical assistance.

\section{Conclusions}
\label{sec:conclusions}
For the above transformation to be realized, it is important that medical institutions drive themselves the computational medical XR knowledge transformation via XR content curriculum simulators and continuous education of their professionals, for the immediate benefit of their patients with significant impact to humanity in general.

Let’s make the world a better place through computational science and science-based tools, by advancing the field and applications of computational medical XR.

\section{Acknowledgements}
\label{sec:acknowledgements}

Parts of this article have been presented in the 2nd Swiss Conference for Telemedicine and Digital Health 2021, Bern, Switzerland as well as a Frontiers Workshop and Talk at ACM SIGGRAPH 2023 \cite{gp2021compXRmed}. Special thanks to Craig Patterson for reviewing this article, and to Dr. Kyriakos Vamvakidis for his support and guidance in the intraoperative neuromonitoring use-case. 
Heartfelt gratitude is extended to George Sofianos, Angeliki Karava and Alexandros Sgouridis for their generous provision of a modern surgical robotic surgery system and invaluable help and support throughout the
learning journey. Lastly, special thanks to Pearly Chen and HTC Vive for providing
us early access to the HTC Vive Ultimate body trackers. This work is partially supported by the OMEN-E project (PFSA22-240), that have received funding from Innosuisse Accelerator programme, 5G-EPICENTRE (GA No 101016521) and FIDAL projects (GA No 101096146), that have received funding from the European Union’s Horizon 2020 and Horizon Europe research and innovation programmes respectively. Additionally, the work in section \ref{sec:IDS} was partially funded by the Innosuisse Impulse programme IDS (100.133 IP-ICT). Finally, the work in section \ref{sec:VOSCE}
was partially funded by the Innosuisse Impulse programme Viva VOSCE (105.072 IP-ICT).

% \clearpage
\bibliographystyle{spmpsci}
\bibliography{references} 

\begin{thebibliography}{10}
\providecommand{\url}[1]{{#1}}
\providecommand{\urlprefix}{URL }
\expandafter\ifx\csname urlstyle\endcsname\relax
  \providecommand{\doi}[1]{DOI~\discretionary{}{}{}#1}\else
  \providecommand{\doi}{DOI~\discretionary{}{}{}\begingroup \urlstyle{rm}\Url}\fi

\bibitem{5G-EPICENTRE}
Project: 5g-epicentre (2021).
\newblock \urlprefix\url{https://www.5gepicentre.eu/}

\bibitem{IDS}
Project: Intelligent digital surgeon (2022).
\newblock \urlprefix\url{https://miralab.com/project/intelligent-digital-surgeon-ids/}

\bibitem{FIDAL}
Project: Fidal (2023).
\newblock \urlprefix\url{https://fidal-he.eu/}

\bibitem{VOSCE}
Project: Vivavosce (2023).
\newblock \urlprefix\url{https://www.hug.ch/en/neurocentre/centre-virtual-medicine}

\bibitem{AccentureNewReality}
Accenture: Waking up to a new reality (2020).
\newblock \urlprefix\url{https://www.accenture.com/_acnmedia/Accenture/Redesign-Assets/DotCom/Documents/Global/1/Accenture-G20-YEA-report.pdf}

\bibitem{bansal2022healthcare}
Bansal, G., Rajgopal, K., Chamola, V., Xiong, Z., Niyato, D.: Healthcare in metaverse: A survey on current metaverse applications in healthcare.
\newblock Ieee Access \textbf{10}, 119,914--119,946 (2022)

\bibitem{Bevan2019}
Bevan, J., Russell, B., Marshall, B.: A new approach to osce preparation—prosces.
\newblock BMC Medical Education \textbf{19}(1), 126 (2019).
\newblock \doi{10.1186/s12909-019-1571-5}

\bibitem{Bloom1986}
Bloom, B.: Taxonomy of educational objectives. 1: Cognitive domain, 29 edn.
\newblock Longman (1986)

\bibitem{cate2023current}
Cate, G., Barnes, J., Cherney, S., Stambough, J., Bumpass, D., Barnes, C.L., Dickinson, K.J.: Current status of virtual reality simulation education for orthopedic residents: the need for a change in focus.
\newblock Global Surgical Education-Journal of the Association for Surgical Education \textbf{2}(1), 46 (2023)

\bibitem{CVM}
CVM: Centre for virtual medicine (2023).
\newblock \urlprefix\url{https://www.hug.ch/en/neurocentre/centre-virtual-medicine}

\bibitem{enhanceHealthcarePostPandemic}
Dzyuba, A.: How immersive technology could enhance healthcare of the post-pandemic reality (2021).
\newblock \urlprefix\url{https://tinyurl.com/hrarczdm}

\bibitem{vrIsleAcademy}
Filippidis, A., Marmaras, N., Maravgakis, M., Plexousaki, A., Kamarianakis, M., Papagiannakis, G.: Vr isle academy: A vr digital twin approach for robotic surgical skill development.
\newblock Accepted for presentation at CGI  (2024)

\bibitem{cologne}
Geronikolakis, S., Kamarianakis, M., Papagiannakis, G.: Enhancing abdominal surgical training through immersive vr simulations with gamified educational elements.
\newblock Work in progress  (2024)

\bibitem{GreenleafVRHealthcare}
Greenleaf, W.: How virtual reality technology will impact healthcare (2021).
\newblock \urlprefix\url{https://www.linkedin.com/pulse/how-virtual-reality-technology-impact-healthcare-greenleaf-phd/}

\bibitem{vrada2021}
Hassandra, M., Galanis, E., Hatzigeorgiadis, A., Goudas, M., Mouzakidis, C., Karathanasi, E.M., Petridou, N., Tsolaki, M., Zikas, P., Evangelou, G., Papagiannakis, G., Bellis, G., Kokkotis, C., Panagiotopoulos, S.R., Giakas, G., Theodorakis, Y.: $\alpha$ virtual reality app for physical and cognitive training of older people with mild cognitive impairment: Mixed methods feasibility study.
\newblock JMIR Serious Games \textbf{9}(1), e24,170 (2021).
\newblock \doi{10.2196/24170}.
\newblock \urlprefix\url{https://games.jmir.org/2021/1/e24170}

\bibitem{herur2021next}
Herur-Raman, A., Almeida, N.D., Greenleaf, W., Williams, D., Karshenas, A., Sherman, J.H.: Next-generation simulation—integrating extended reality technology into medical education.
\newblock Frontiers in Virtual Reality \textbf{2}, 693,399 (2021)

\bibitem{HOOPER2019}
Hooper, J., Tsiridis, E., Feng, J.E., Schwarzkopf, R., Waren, D., Long, W.J., Poultsides, L., Macaulay, W., Papagiannakis, G., Kenanidis, E., Rodriguez, E.D., Slover, J., Egol, K.A., Phillips, D.P., Friedlander, S., Collins, M.: Virtual reality simulation facilitates resident training in total hip arthroplasty: A randomized controlled trial.
\newblock The Journal of Arthroplasty \textbf{34}(10), 2278--2283 (2019).
\newblock \doi{https://doi.org/10.1016/j.arth.2019.04.002}.
\newblock \urlprefix\url{https://www.sciencedirect.com/science/article/pii/S0883540319303341}

\bibitem{ICRCvr}
ICRC: Virtual reality innovation (2021).
\newblock \urlprefix\url{https://www.icrc.org/en/what-we-do/virtual-reality}

\bibitem{InsiderIntelligence}
Intelligence, I.: What types of new applications/solutions do us ar/vr professionals expect immersive technologies to offer in the healthcare sector? (2020).
\newblock \urlprefix\url{https://tinyurl.com/sc8az5wh}

\bibitem{NeverDropTheBall2021}
Kamarianakis, M., Lydatakis, N., Papagiannakis, G.: Never ‘drop the ball’in the operating room: An efficient hand-based vr hmd controller interpolation algorithm, for collaborative, networked virtual environments.
\newblock In: Computer Graphics International Conference, pp. 694--704. Springer (2021)

\bibitem{kamarianakis2023ar}
Kamarianakis, M., Protopsaltis, A., Papagiannakis, G.: Ar-assisted surgical care via 5g networks for first aid responders.
\newblock arXiv preprint arXiv:2303.00458  (2023)

\bibitem{ireactPoster2024}
Kentros, M., Filippidis, A., Petropoulos, J., Cole, M., Popov, V., Harmer, B., Kamarianakis, M., Papagiannakis, G.: Enhancing crew resource management in vr medical training: Tools for analyzing advanced, collaborative group training in the metaverse.
\newblock Submitted to SIGGRAPH  (2024)

\bibitem{Khan2013PartI}
Khan, K.Z., Ramachandran, S., Gaunt, K., Pushkar, P.: The objective structured clinical examination (osce): Amee guide no. 81. part i: An historical and theoretical perspective.
\newblock Medical Teacher \textbf{35}(9), e1437--e1446 (2013).
\newblock \doi{10.3109/0142159X.2013.818634}

\bibitem{transG}
Ma, L., Kang, H., Thalmann, N.M., Wac, K.: Transg: a spatial-temporal transformer for surgical gesture recognition.
\newblock Accepted for presentation at CGI  (2024)

\bibitem{gp2021compXRmed}
Papagiannakis, G., Greenleaf, W.: Frontiers talk: Computational medical xr: Spatial, neural and wearable computing converging to transform healthcare (2023).
\newblock \urlprefix\url{https://s2023.siggraph.org/presentation/?id=ftalk_101&sess=sess408}

\bibitem{frontiersUpskillResearchTopic}
Papagiannakis, G., Kannape, O., Greenleaf, W., Cole, M., Jones, G., Zhang, M., Herbelin, B.: Frontiers workshop: Computational medical xr (2023).
\newblock \urlprefix\url{https://s2023.siggraph.org/presentation/?id=fwork_109&sess=sess287}

\bibitem{Just2003}
Ponder, M., Herbelin, B., Molet, T., Schertenlieb, S., Ulicny, B., Papagiannakis, G., Magnenat-Thalmann, N., Thalmann, D.: Immersive vr decision training: Telling interactive stories featuring advanced virtual human simulation technologies.
\newblock In: Proceedings of the Workshop on Virtual Environments 2003, EGVE '03, p. 97–106. Association for Computing Machinery, New York, NY, USA (2003).
\newblock \doi{10.1145/769953.769965}.
\newblock \urlprefix\url{https://doi.org/10.1145/769953.769965}

\bibitem{GrandViewResearch}
Research, G.: Virtual reality market size, share trends analysis report by technology (2021).
\newblock \urlprefix\url{https://www.grandviewresearch.com/industry-analysis/virtual-reality-vr-market}

\bibitem{XRukHealthcareReport}
Research, U., Innovation: The growing value of xr in healthcare in the united kingdom (2021).
\newblock \urlprefix\url{https://www.xrhealthuk.org/the-growing-value-of-xr-in-healthcare#Contents}

\bibitem{RivaEmbodiedSimulation}
Riva, G., Wiederhold, B., Mantovani, F.: Neuroscience of virtual reality: From virtual exposure to embodied medicine.
\newblock Cyberpsychology, Behavior, and Social Networking \textbf{22}, 82--96 (2019).
\newblock \doi{10.1089/cyber.2017.29099.gri}

\bibitem{vrForTraining}
Slater, M.: Virtual reality for training (2020).
\newblock \urlprefix\url{http://presence-thoughts.blogspot.com/2020/03/virtual-reality-for-training.html}

\bibitem{10.3389/frobt.2016.00074}
Slater, M., Sanchez-Vives, M.V.: Enhancing our lives with immersive virtual reality.
\newblock Frontiers in Robotics and AI \textbf{3}, 74 (2016).
\newblock \doi{10.3389/frobt.2016.00074}.
\newblock \urlprefix\url{https://www.frontiersin.org/article/10.3389/frobt.2016.00074}

\bibitem{medicalXRtaxonomy}
Spiegel, B.M., Rizzo, A., Persky, S., Liran, O., Wiederhold, B., Woods, S., Donovan, K., Sarkar, K., Xiang, H., Joo, S., Jotwani, R., Lang, M., Paul, M., Senter-Zapata, M., Widmeier, K., Zhang, H.: What is medical extended reality? a taxonomy defining the current breadth and depth of an evolving field.
\newblock Journal of Medical Extended Reality \textbf{1}(1), 4--12 (2024).
\newblock \doi{10.1089/jmxr.2023.0012}.
\newblock \urlprefix\url{https://doi.org/10.1089/jmxr.2023.0012}

\bibitem{MedicalProcedures}
wikipedia: Medical procedure (2021).
\newblock \urlprefix\url{https://en.wikipedia.org/wiki/Medical_procedure}

\bibitem{cvrsb}
Zikas, P., Kamarianakis, M., Kartsonaki, I., Lydatakis, N., Kateros, S., Kentros, M., Geronikolakis, E., Evangelou, G., Apostolou, A., Catilo, P.A.A., et~al.: Covid-19-vr strikes back: innovative medical vr training.
\newblock In: ACM SIGGRAPH 2021 Immersive Pavilion, pp. 1--2 (2021)

\bibitem{Zikas2020VRpatterns}
Zikas, P., Papagiannakis, G., Lydatakis, N., Kateros, S., Ntoa, S., Adami, I., Stephanidis, C.: Immersive visual scripting based on {VR} software design patterns for experiential training.
\newblock Vis. Comput. \textbf{36}(10), 1965--1977 (2020).
\newblock \doi{10.1007/s00371-020-01919-0}.
\newblock \urlprefix\url{https://doi.org/10.1007/s00371-020-01919-0}

\bibitem{10038619}
Zikas, P., Protopsaltis, A., Lydatakis, N., Kentros, M., Geronikolakis, S., Kateros, S., Kamarianakis, M., Evangelou, G., Filippidis, A., Grigoriou, E., Angelis, D., Tamiolakis, M., Dodis, M., Kokiadis, G., Petropoulos, J., Pateraki, M., Papagiannakis, G.: Mages 4.0: Accelerating the world’s transition to vr training and democratizing the authoring of the medical metaverse.
\newblock IEEE Computer Graphics and Applications \textbf{43}(2), 43--56 (2023).
\newblock \doi{10.1109/MCG.2023.3242686}

\end{thebibliography}

\end{document}